Appl.Math.Lett 8,N4,(1995),87-90.
\input amstex
\documentstyle{amsppt}

\hsize32pc
\vsize50pc

\hcorrection{.5in}

\TagsOnRight

\topmatter
\title Examples of nonuniqueness for an inverse \\
       problem of geophysics \footnotemark "*"\endtitle
\footnotetext "*" {This work was done under the auspices
of DOE. \hfill}
\rightheadtext{Examples}
\author A.G. Ramm \endauthor
\affil  Department of Mathematics, Kansas State University,
        Manhattan, KS  66506-2602, USA \endaffil

\abstract {Two different velocity profiles and a source term are
constructed such that the surface data are the same 
for all times and are not identically zero.
The
governing equation is  $c^{-2} (x) u_{tt} - \Delta u = f(x,t)$  
in  $D
\times [0,\infty)$, $u = 0$  for  $t < 0$, $u_N = 0$  on  $\partial D$,
$D \subset {\Bbb R}_+^n := \{ x : x_n > 0\}$,
$f(x,t) \not\equiv 0$.  The data are the values
$u(x,t)$, $\forall x \in S$, $\forall t > 0$.  Here  $S$  is the part of
$\partial D$  which lies on the plane  $x_n = 0$, $D = \{ x : a_j \leq
x_j \leq b_j, 1 \leq j \leq n, a_n = 0\}$.} \endabstract

\endtopmatter

\vglue .2in

\document

\subhead I. Introduction \endsubhead

Let  $D \subset {\Bbb R}_+^n := \{ x : x \in {\Bbb R}^n, x_n \geq 0\}$
be a bounded domain, part  $S$  of the boundary  $\Gamma$  of  $D$  is
on the plane  $x_n = 0$, $f(x,t)$  is a source of the wavefield, $c(x) >
0$  is a velocity profile.  The wavefield, e.g., the acoustic pressure,
solves the problem:
$$
c^{-2} (x) u_{tt} - \Delta u = f(x,t) \quad\text{in}\quad D \times [0,
\infty),\quad f(x,t) \not\equiv 0, \tag 1
$$
$$
u_N = 0 \quad\text{on}\quad \Gamma \tag 2
$$
$$
u = u_t = 0 \quad\text{at}\quad t = 0.  \tag 3
$$
Here  $N$  is the unit outer normal to  $\Gamma$, $u_N$  is the normal
derivative of  $u$  on  $\Gamma$.  If  $c^2 (x)$  is known, then the
direct problem (1)-(3) is uniquely solvable.  The inverse problem (IP)
we are interested in is the following one:

\vskip .2in

\noindent (IP) {\it Given the data  $u(x,t) \quad \forall x \in S$,
$\forall t > 0$, can one recover  $c^2 (x)$  uniquely?}

\vskip .2in

The basic result of this paper is: {\it the answer to the above question
is no}. 

 An analytical construction is presented of two constant
velocities  $c_j > 0$, $c_1 \not= c_2$, which can be chosen arbitrary,
and a source, which is constructed after  $c_j > 0$  are chosen, such
that the solutions to problems (1)-(3) with  $c^2 (x) = c_j^2$  produce
the same surface data on  $S$  for all times:
$$
u_1 (x,t) = u_2 (x,t) \quad \forall x \in S, \quad \forall t > 0.
\tag 4
$$
The domain  $D$  we use is a box: $D = \{ x : a_j \leq x_j \leq b_j, 1
\leq j \leq n\}$.

This construction is given in section II. At the end of section II the
data on  $S$  are suggested, which allow one to uniquely determine  $c^2
(x)$.

\vskip .2in

\subhead II. Example of nonuniqueness of the solution to IP \endsubhead

Our construction is valid for any  $n \geq 2$.  For simplicity we take
$n = 2$, $D = \{ x : 0 \leq x_1 \leq \pi, 0 \leq x_2 \leq \pi\}$.  Let
$c^2 (x) = c^2 = \text{const} > 0$.  The solution to (1)-(3) with  $c^2
(x) = c^2 = \text{const}$  can be found analytically
$$
u(x,t) = \sum_{m=0}^\infty u_m (t) \phi_m (x), \quad m = (m_1,m_2)
\tag 5
$$
where
$$
\aligned
\phi_m (x) &= \gamma_{m_1m_2} \cos (m_1 x_1) \cos (m_2 x_2), \\
\int_D \phi_m^2(x)dx &= 1, \quad \Delta \phi_m +\lambda_m \phi_m = 0, \\
\phi_{m N}&= 0 \quad\text{on}\quad \Gamma, \quad \lambda_m := m_1^2 +m_2^2,
\\
\gamma_{00} &= \frac 1\pi, \quad \gamma_{m_10} = \gamma_{0m_2} = \frac
{\sqrt{2}}\pi, \endaligned\tag 6
$$
$$
\gamma_{m_1m_2} = 2/\pi \quad\text{if}\quad m_1 > 0 \quad\text{and}\quad
m_2 > 0,
$$
$$
u_m (t) := u_m (t,c) = \frac c{\sqrt{\lambda_m}} \int_0^t \sin [c
\sqrt{\lambda_m} (t - \tau)] f_m (\tau) d\tau, \quad f_m (t) := \int_D
f(x,t) \phi_m (x) dx \tag 6$^\prime$
$$
  The data are
$$
u(x_1,0,t) = \sum_{m=0}^\infty u_m (t,c) \gamma_{m_1 m_2} \cos (m_1
x_1). \tag 7
$$
For these data to be the same for  $c = c_1$  and  $c= c_2$, it is
necessary and sufficient that
$$
\sum_{m_2 = 0}^\infty \gamma_{m_1 m_2} u_m (t,c_1) =
\sum_{m_2 = 0}^\infty \gamma_ {m_1 m_2}u_m (t,c_2), \quad \forall t >
0,\quad \forall m_1.  \tag 8
$$
Taking Laplace transform of (8) and using (6$^\prime$) one gets an
equation, equivalent to (8),
$$
\sum_{m_2=0}^\infty
\gamma_ {m_1 m_2} \overline{f}_m (p) \left[ \frac {c_1^2}{p^2 + c_1^2
\lambda_m} - \frac {c_2^2}{p^2 + c_2^2 \lambda_m}\right] = 0, \quad
\forall p > 0,\quad \forall m_1.  \tag 9
$$

Take  $c_1 \not= c_2$, $c_1, c_2 > 0$, arbitrary and find
$\overline{f}_m (p)$  for which (9) holds.  This can be done by
infinitely many ways.  Since (9) is equivalent to (8), the desired
example of nonuniqueness of the solution to IP is constructed.

Let us give a specific choice: $c_1 = 1$, $c_2 = 2$,
$\overline{f}_{m_1m_2} = 0$  for  $m_1 \not= 0$, $m_2 \not= 1$  or  $m_2
\not= 2$, $\overline{f}_{02} (p) = \frac 1{p + 1}$, $\overline{f}_{01}
(p) = -\frac {p^2 + 1}{(p + 1)(p^2 + 16)}$.  Then (9) holds.  Therefore,
if
$$
f(x,t) = \frac {\sqrt{2}}\pi \left[ f_{01} (t) \cos (x_2) + f_{02} (t)
\cos (2x_2)\right], \quad c_1 = 1, \quad c_2 = 2 \tag 10
$$
then the data  $u_1 (x,t) = u_2 (x,t) \quad \forall x \in S$, $\forall t
> 0$.  In (10) the values of the coefficients are
$$
f_{01} (t) = -\frac 2{17} \exp (-t) - \frac {15}{17} \left[ \cos (4t) -
\frac 14 \sin (4t)\right], \quad f_{02} (t) = \exp (-t).  \tag 11
$$

\vskip .2in

\demo{Remark 1} The above example brings out the question: What data on
$S$  are sufficient for the unique identifiability of  $c^2 (x)$?  The
answer to this question one can find in [1] and [2].

In particular, if one takes  $f(x,t) = \delta (t) \delta (x - y)$, and
allows  $x$  and  $y$  run through  $S$, then the data  $u(x,y,t) \quad
\forall x,y \in S$, $\forall t > 0$, determine  $c^2 (x)$  uniquely.  In
fact, the low frequency surface data  $\tilde u(x,y,k)$, $\forall x,y
\in S \quad \forall k \in (0,k_0)$, where  $k_0 > 0$  is an arbitrary
small fixed number, determine  $c^2 (x)$  uniquely under mild
assumptions on  $D$  and  $c^2 (x)$.  By  $\tilde u (x,y,k)$  is meant
the Fourier transform of  $u(x,y,t)$  with respect to  $t$.
\enddemo

\demo{Remark 2} One can check that the non-uniqueness example with
constant velocities is not possible to construct as was done above
if the sources are concentrated on $S$, that is, if $f(x_1,x_2,t)=
\delta (x_2)f_1(x_1,t)$.
\enddemo

\vskip .2in

\noindent {\bf Acknowledgement.} The author thanks Prof.~S.~Seatzu who
drew his attention to the IP discussed above.

\vfill

e-mail: RAMM\@MATH.KSU.EDU

\enddocument